\documentclass[prb,preprint,showpacs,preprintnumbers,amsmath,amssymb]{revtex4}
\usepackage{graphicx}% Include figure files

\usepackage{dcolumn}% Align table columns on decimal point
\usepackage{bm}% bold math
\usepackage[mathscr]{eucal}
\usepackage{mathrsfs}

\begin{document}

% Full title of the paper (Capitalized)
% \title{Specific heat jump of silica glass}
% \title{First-principles calculation of the specific heat jump in the glass transition of silica glass}
% \title{First-principles calculation of the glass transition of silica glass and its specific heat jump}
\title{First-principles study on the specific heat jump in the glass transition of silica glass and the Prigogine-Defay ratio}

% Authors, for the paper (add full first names)
\author{Koun Shirai, Kota Watanabe, and Hiroyoshi Momida}
% \email{koun@sanken.osaka-u.ac.jp}
\affiliation{%
The Institute of Scientific and Industrial Research, Osaka University, 8-1 Mihogaoka, Ibaraki, Osaka 567-0047, Japan
}%

\author{Sangil Hyun}
\affiliation{Korea Institute of Ceramic Engineering and Technology,
101 Soho-ro, Jinju-si, Gyeongsangnam-do, 52851, Korea}

% Abstract (Do not use inserted blank lines, i.e. \\) 
\begin{abstract}
The most important characteristic of glass transition is a jump in specific heat $\Delta C_{p}$. Despite its significance, no standard theory exists to describe it. In this study, first-principles molecular-dynamics (MD) simulations are used to describe the glass transition of silica glass, which presents many challenges.
The novel view that state variables are extended to include the equilibrium positions of atoms $\{ \bar{\bf R}_{j} \}$ is fully used in analyzing the simulation results.
Decomposing the internal energy into three components (structural, phonon, and thermal expansion energies) reveals that the jump $\Delta C_{p}$ of silica glass is entirely determined by the component of structural energy. The reason for the small $\Delta C_{p}$ is its high glass-transition temperature, which makes the fluctuation in the structural energy insensitive to temperature changes.
This significantly affects how the Prigogine-Defay ratio $\it{\Pi}$ is interpreted, which was previously unknown. The ratio $\it{\Pi}$ represents the ratio of the total energy change to the contribution of thermal expansion energy at the glass transition. The general property, $\it{\Pi}>1$, of glasses indicates that glass transitions occur mainly by changes in the structural energy. Silica glass is an extreme case in that the transition occurs entirely by the change in internal structure, such as the distribution of the bending angle of Si--O--Si bond.
\end{abstract}

\pacs{Rev. 1.4, (8 Feb 2022)}
% deflect from

\maketitle
%%%%%%%%%%%%%%%%%%%%%%%%%%%%%%%%%%%%%%%%%%
%% Only for the journal Gels: Please place the Experimental Section after the Conclusions

%%%%%%%%%%%%%%%%%%%%%%%%%%%%%%%%%%%%%%%%%%

\section{Introduction}
\label{sec:intro}
Silica glass (SiO$_{2}$) is an archetypal glass with the simplest chemical formula. However, it could be the least understood of all glasses, as it is expressed by the word  ``a deceitful simplicity'' (Chap.~5 of Ref.~\onlinecite{Mysen-Richet05}). It exhibits several unusual properties, including its extremely large Prigogine-Defay ratio (larger than the usual by four-order of magnitude) \cite{Nemilov-VitreousState} and its exceptional disparity between the thermodynamic and dynamic fragilities \cite{Martinez01,Richet84a}. Its thermal expansivity, $\alpha$, is exceptionally small, whereas its compressibility, $\kappa$, is normal.
Further, its thermal expansion at high pressure is unusually large.\cite{Yang20} These, as well as other properties, are yet to be resolved.
% \cite{Schmelzer06}

Experimentally, a limitation to studying the thermodynamic properties of silica glass is its large viscosity, which prevents the glass from achieving equilibrium. The contrast between the liquid and glass is relatively easy to discern in other glasses by measuring viscosity $\eta$; the temperature at which $\eta$ reaches about $10^{13}$ Poise agrees with the calorimetric transition temperature, $T_{g}$, obtained by specific-heat measurement. Alternatively, the viscosity of silica glass varies only gradually over a wide temperature, $T$, range; no discernible characteristic temperature exists. Practically, the glass manufacturing process uses three characteristic temperatures: strain ($T_{\rm st}$), annealing ($T_{\rm an}$), and softening temperatures ($T_{\rm sf}$);\cite{Sudoh09} this fact alone indicates that there is no sharp transition. 
This could explain why the viscosity of silica glass obeys the Arrhenius law.

\begin{figure}[htbp]
    \centering
    \includegraphics[width=100 mm, bb=0 0 420 280]{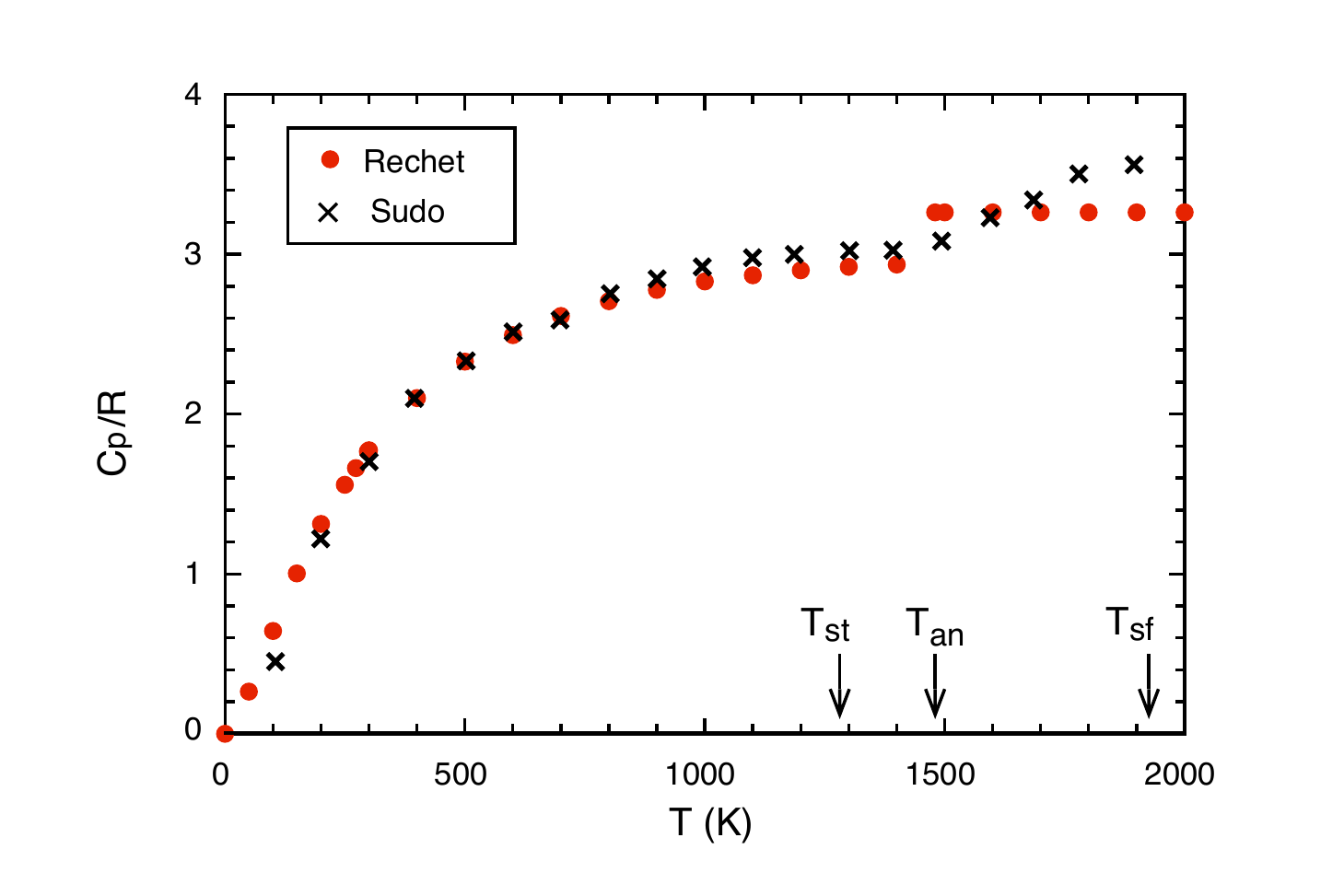} 
  \caption{Specific heat of silica glass: red circles indicate Richet {\it et al.}’s data obtained using the drop calorimetric method \cite{Richet82}; crosses indicate Sudo’s data obtained using the adiabatic calorimetric method \cite{Sudoh99}. 
  The three characteristic temperatures in the viscosity measurements are indicated at the bottom: the strain ($T_{\rm st}$ defined at log $\eta$(poise) = 14.5), annealing ($T_{\rm an}$: log $\eta=$ 13), and softening ($T_{\rm sf}$: log $\eta=$ 6.5) temperatures.
  } \label{fig:CTexperiment}
\end{figure}

The best method, notwithstanding existing objections, is the specific-heat measurement. Isobaric specific heat exhibits a jump of $\Delta C_{p} = C_{p}^{(l)}-C_{p}^{(g)}$, where $C_{p}^{(l)}$ and $C_{p}^{(g)}$ are the specific heat of the liquid and glass states, respectively. Using this method, the glass-transition temperature, $T_{g}$, of silica glass is generally accepted as $T_{g}=1480$ K.\cite{Bruckner71} Figure \ref{fig:CTexperiment} shows the experimental data for $C_{p}$ as a function of $T$, compared with the three characteristic temperatures in the viscosity measurement.\cite{Sudoh99,Richet82} The figure shows consistency in $T_{g}$ between the two data; however, Br\"{u}ckner noted in his review paper that there are data missing this jump.\cite{Bruckner71} Although the two data are consistent at the starting temperature of change in the slope in the $C_{p}$-$T$ curve, the magnitude, $\Delta C_{p}$, in Sudo's data is unclear due to the continuous increase in $C_{p}^{(l)}$. However, we can accept the value $\Delta C_{p}=8.0$ J/mol$\cdot$K (or $0.32 R$ per mole of atoms) for the jump $\Delta C_{p}$, which is about 10\% of the total specific heat at $T_{g}$.\cite{Richet90} The gas constant, $R$, is used as per mole of average atoms throughout this study, indicating that the classical limit of isochoric specific heat, $C_{v}$, is $3R$.
This $\Delta C_{p}$ value of silica glass may be the smallest among other glasses. It is said that a correlation between the jump in $\Delta C_{p}$ and the glass fragility exists. This is a significant challenge in the current glass research.\cite{Angell95,Takahara95,Ngai99,Johari00,Xia00,Martinez01,Granato02,Lubchenko07, Garrahan03,Biroli05,Chandler10}
Although the jump implies an abrupt change, the glass-transition temperature has a finite width, $\Delta T_{g}$, which is called {\it the transition region}.\cite{Moynihan74} In Sudo's data, the width is unclear, whereas Richet {\it et al}’s data show an abrupt change of less than 100 K in the width. Richet and Bottinga investigated $C_{p}$ in the transition region in terms of the fictive temperature,\cite{Richet84b} from which they obtained the data shown in Fig.~\ref{fig:CTexperiment}. 
In addition, the specific heat of the liquid $C_{p}^{(l)}(T)$ differs for the aforementioned two data: one is nearly constant with respect to $T$, whereas the other increases; the increasing dependence is also shown in the review paper by Br\"{u}ckner \cite{Bruckner71}. The last two issues with $\Delta T_{g}$ and $C_{p}^{(l)}(T)$ receive little consideration in this study.

Theoretically, a nonempirical study on $\Delta C_{p}$ and $T_{g}$ for silica glass is absent, primarily due to a lack of rigorous theory of specific heat for liquids.
Instead, various models for the glass transition of oxide glasses were proposed, which are reviewed by Ojovan.\cite{Ojovan08} There is no convincing answer to the question of why the jump $\Delta C_{p}$ of silica glass is so small. A recent study by Trachenko and Brazhkin calculated $\Delta C_{p}$ using an elastic model, and the result was consistent with experiments.\cite{Trachenko11} This model may convey some of real properties of glasses. However, as will be discussed later, consistency with other properties is important for understanding the glass transition, thus a method without empirical parameters is required.
This is the first purpose of this study. First-principles molecular dynamics (FP-MD) simulations are used to investigate the glass transition of silica glass by calculating thermodynamic properties. 
Particular attention is given to the specific-heat jump, due to the aforementioned importance of $\Delta C_{p}$ in the glass transition. 
When studying the mechanism that determines $\Delta C_{p}$, it is critical to specify which state variables determine the internal energy of solids. This is the subject of the Prigogine-Defay (PD) ratio, which has yet to be assigned a concrete interpretation. In this regard, the novel view on the state variables of solids, which was proposed by the first author of this paper,\cite{Shirai20-GlassState,StateVariable} has significant merit. Based on this view, we can reveal the determining mechanism for $\Delta C_{p}$ at a microscopic level. The second purpose of this study is to describe this concern.

The paper is organized as follows. A rigorous theory of specific heat for both solids and liquids is provided in Sec.~\ref{sec:Backgrounds}. A detailed explanation of the theory is provided elsewhere.\cite{Shirai22-SH} The theory’s essential components are briefly described. The results of applying the theory to silica glass are described in Sec.~\ref{sec:results}. Although we are mainly concerned with the behavior of the specific heat, structural analysis is also included. This is because it is found that the PD ratio is determined by details of the structural change in the glass transition. The relation between the structural change and the PD ratio is discussed in Sec.~\ref{sec:discussion}. The last section shows the conclusion of this study.

%%%%%%%%%%%%%%%%%%%%%%%%%%%%%%%%%%%%%%%%%%%%%%%%
% \section{Methodologies}
\section{Calculation method of specific heat}  % of glasses and liquids
\label{sec:Backgrounds}

\subsection{Theory}
\label{sec:method}

\paragraph{Specific heat.}
The study of glass transitions requires a thorough formulation of specific heat for liquids, which has only lately been established.\cite{Shirai22-SH} The original paper \onlinecite{Shirai22-SH} should be consulted for more information.
The specific heat is defined as the change in internal energy, $U$, with respect to a small change in temperature $T$. The internal energy of a material is defined at equilibrium and is given by the time average of the microscopic total energy, $E_{\rm tot}( \{ {\bf R}_{j}(t) \} )$, of the material; where ${\bf R}_{j}$ is the position of $j$th atom, while there are $N$ atoms in the material. According to density functional theory (DFT), the total energy, $E_{\rm tot}$, is the sum of the ground-state energy, $E_{\rm gs}( \{ {\bf R}_{j}(t) \} )$, and the kinetic energy of the atoms, $E_{\rm K} = (1/2) \sum_{j} M_{j} v_{j}(t)^{2}$ ($M_{j}$ and $v_{j}$ are mass and velocity, respectively, of $j$th atom), and is expressed as follows:
\begin{equation}
U = \overline{E_{\rm tot}(t)} = \overline{E_{\rm gs}( \{ {\bf R}_{j}(t) \})}
    + \frac{1}{2} \sum_{j} M_{j} \overline{ v_{j}(t)^{2} }.
\label{eq:internal-energy}
\end{equation}
We can determine the relationship between $U$ and $T$ by performing MD runs as either $U$ or $T$ varies, with the volume, $V$, fixed, and deducing isochoric specific heat $C_{v} = (\partial U/\partial T)_{v} $. The formulation is quite generic up to this point. Because the total energy is well defined by DFT, $C_{v}$ can be directly calculated by DFT-based MD simulations, irrespective of whether a material is solid or liquid. 

The harmonic approximation holds for solids at low temperatures. The instantaneous position of an atom is the sum of the equilibrium position, $\bar{\bf R}$, and small displacement, $\bar{\bf u}$, from it, and is expressed as ${\bf R}_{j}(t) = \bar{\bf R}_{j} + {\bf u}_{j}(t)$. The ground-state energy, $E_{\rm gs}( \{ {\bf R}_{j}(t) \} )$, can be expanded with respect to the displacement, a part of this energy becomes the phonon energy, $E_{\rm ph} = \sum_{q} (\bar{n}_{q} + 1/2 ) \hbar \omega_{q}$ ($\omega_{q}$ is the frequency of $q$th phonon and $\bar{n}_{q}$ is its Bose occupation number). Thus, the internal energy, $U$, can be decomposed as follows:
\begin{equation}
U \equiv U(T,V,\{ \bar{\bf R}_{j} \} ) = E_{\rm st}( \{ \bar{\bf R}_{j} \} )  +  E_{\rm ph}(T) + E_{\rm te}(V).
\label{eq:internal-energy-3}
\end{equation}
In this paper, {\em the structural energy} $E_{\rm st}$ is defined as $E_{\rm st}=\overline{E_{\rm tot}(t)} -E_{\rm ph} = \overline{E_{\rm gs}(t)}-(1/2) E_{\rm ph}$. While $\{ \bar{\bf R}_{j} \}$ initially denoted a set of $N$ variables, six variables representing the shape of the unit cell has been subtracted from the arguments, $\{ \bar{\bf R}_{j} \}$ of $E_{\rm st}$, in Eq.~(\ref{eq:internal-energy-3}). Therefore, the structural changes in this paper mean the changes in the internal (relative) coordinates with fixing the lattice parameters. The remaining six variables constitute the last term of Eq.~(\ref{eq:internal-energy-3}), which is {\em the thermal expansion energy}, $E_{\rm te}$, for isotropic materials. Although, in a strict sense, these three energies, $E_{\rm st}$, $E_{\rm ph}$, and $E_{\rm te}$, are dependent on all the variables $T$, $V$, and $\{ \bar{\bf R}_{j} \}$, the variable with the most influence on the respective terms is indicated in the arguments.

Based on the decomposition (\ref{eq:internal-energy-3}), the total specific heat, $C_{p}$, is expressed as follows:
\begin{equation}
C_{p} =  C_{\rm st}( \{ \bar{\bf R}_{j} \} )  +  C_{\rm ph}(T) + C_{\rm te}(V).
\label{eq:part-SH}
\end{equation}
$E_{\rm st}$ has no temperature dependence when the potential has a harmonic form, and its contribution to the specific heat, $C_{\rm st}$, vanishes. $C_{v}$ is determined entirely by the phonon part: $C_{v}=C_{\rm ph}$.
The contribution of the thermal expansion energy $\Delta C_{\rm te}$ is expressed as follows:\cite{Callen} 
\begin{equation}
  C_{\rm te} =\frac{T V}{\kappa} \alpha^{2}.
\label{eq:Cthermal-exp}
\end{equation}
In this study, two components, $C_{\rm st}$ and $C_{\rm ph}$, are obtained via MD simulation with a constant $V$, whereas $C_{\rm te}$ is calculated from the experimental data of $\alpha$ and $\kappa$.

The following comments should be noted: First, for liquids, only the total energy, $E_{\rm tot}$, and $C_{p}$ (or $C_{v}$) have physical reality. Phonons are not real substances for liquids. 
However, the component $C_{\rm ph}$ can be operationally obtained from frequency spectra of time-dependent velocities and appears to be effective. We can define the phonon contribution, $C_{\rm ph}$, by applying the Bose--Einstein statistics to the frequency spectrum. In this case, $C_{\rm ph}$ should be interpreted as a virtual quantity {\it defined} by the analogy to the solid case.
Second, despite its practical utility, we should not overlook the virtual nature of the phonon picture, particularly its negligence for energy dissipation. This problem is particularly severe when the transition region is considered. This problem is solved by employing adiabatic MD simulations, which automatically provide the correct relationship between $U$ and $T$. 
Third, for liquids, a set of variables, $\{ \bar{\bf R}_{j} \}$, loses its meaning when used as arguments in Eqs.~(\ref{eq:internal-energy-3}) and (\ref{eq:part-SH}). The equilibrium positions of atoms in liquids are underminate. Therefore, $U$ is a function of $V$ and $T$ only. We will show in Sec.~\ref{sec:discussion}  that this property of variables has a significant impact on thermodynamic relationships.

\paragraph{Adiabatic MD simulations.}
It is impossible to trace the entire glass transition process using computer simulations even with classical MD simulations. Our idea is to simulate glass transition by a series of adiabatic MD runs, each of which a simulation is continued until reaching equilibrium. The internal energy, $U$, is gradually changed from run to run by adjusting the input velocities of atoms. The equilibrium positions of the previous run are used for the initial atom positions of the $k$th run: $ {\bf R}_{j}^{(k)}(0) =\bar{\bf R}_{j}^{(k-1)} $, where ${\bf R}_{j}^{(k)}$ are the atom positions of the $k$th run.
Because an exact value of $T$ is unpredictable in adiabatic MD, the intended temperature of each run was obtained via trial and error. 
The rate of cooling/heating is irrelevant in this simulation because there is no way of knowing how much time is expanded between successive runs. However, a temperature interval, $\Delta T^{(k)}$, between subsequent runs mimics a fast and slow rate of temperature change. Thus, {\it fast} and {\it slow} rates are used in this scenario.
A pressure reserver is also avoided for the same reason.
% As previously stated, executing adiabatic MD is crucial; nevertheless, volume reserve, which entails the imposition of a fixed $V$, should also be avoided. 

The time evolution of the averaged displacements, $\overline{x_{j}(t)^{2} }$, is used to determine whether equilibrium is reached. It is considered an equilibrium state of the solid phase when $\overline{x_{j}(t)^{2} }$ shows constant behavior with respect to $t$. However, it is considered an equilibrium state of the liquid phase when linearity between $\overline{x_{j}(t)^{2} }$ and $t$ is observed over the entire simulation time, $t_{\rm SM}$. The diffusion coefficient, $D$, is determined by the slope of the linear relationship.

\subsection{Calculation method}
\label{sec:condition}
The code used for FP-MD simulations is Phase/0,\cite{PHASE} which is a pseudopotential method. The oxygen atom is treated with an ultrasoft potential,\cite{Vanderbilt90} whereas the silicon atom is treated with a norm-conserved potential.\cite{TM91} The electron-correlation potential is calculated using the generalized gradient approximation of the Perdew--Burke--Ernzerhof type \cite{PBE96}.
The cutoff energy of the plane-wave expansion is 30 Ry. One-point ($R$ point) $k$ mesh is used. Time steps from 0.72 to 1.2 fs are used in MD simulations, depending on the convergence. The total simulation time, $t_{\rm SM}$, varies from 2 to 10 ps. When the presence/absence of diffusion is evident, 2 ps was sufficient, whereas long $t_{\rm SM}$ was used for the marginal case.

Structural models for the silica glass were prepared using crystal $\alpha$-quartz. The lattice system of $\alpha$-quartz is hexagonal and the unit-cell lattice parameters are $a = 9.827$, $b= 9.827$, and $c= 10.809$ \AA.
A supercell with dimensions of $2 \times 2 \times 2$ was used. Two different cell sizes were prepared. One is the {\em high-density cell}, which has the original size of the lattice parameters, whereas the other is obtained by slightly expanding the lattice parameters to have the same density as that of silica glass, which is smaller than the former by 5.4\%. It is called the {\em low-density cell}.
The low-density cell was mostly used, except for calculating the structural parameters during crystal melting.

%%%%%%%%%%%%%%%%%%%%%%%%%%%%%%%%%%%%%%%%%%%%%%%%
\section{Results of MD simulations}
\label{sec:results}

\subsection{Melting of crystal}
\label{sec:heating}
In the beginning, the silica liquid was prepared by starting with $\alpha$-quartz crystals.
The goal of this study is not to investigate the crystal melting process. However, understanding this process provides the basis for further analysis on the glass transition; consequently, we describe it with this utility in mind.
The low-density cell was chosen because of its ease of melting. However, it was not as different as was expected.

\begin{figure}[htbp]
    \centering
    \includegraphics[width=120mm, bb=0 0 450 350]{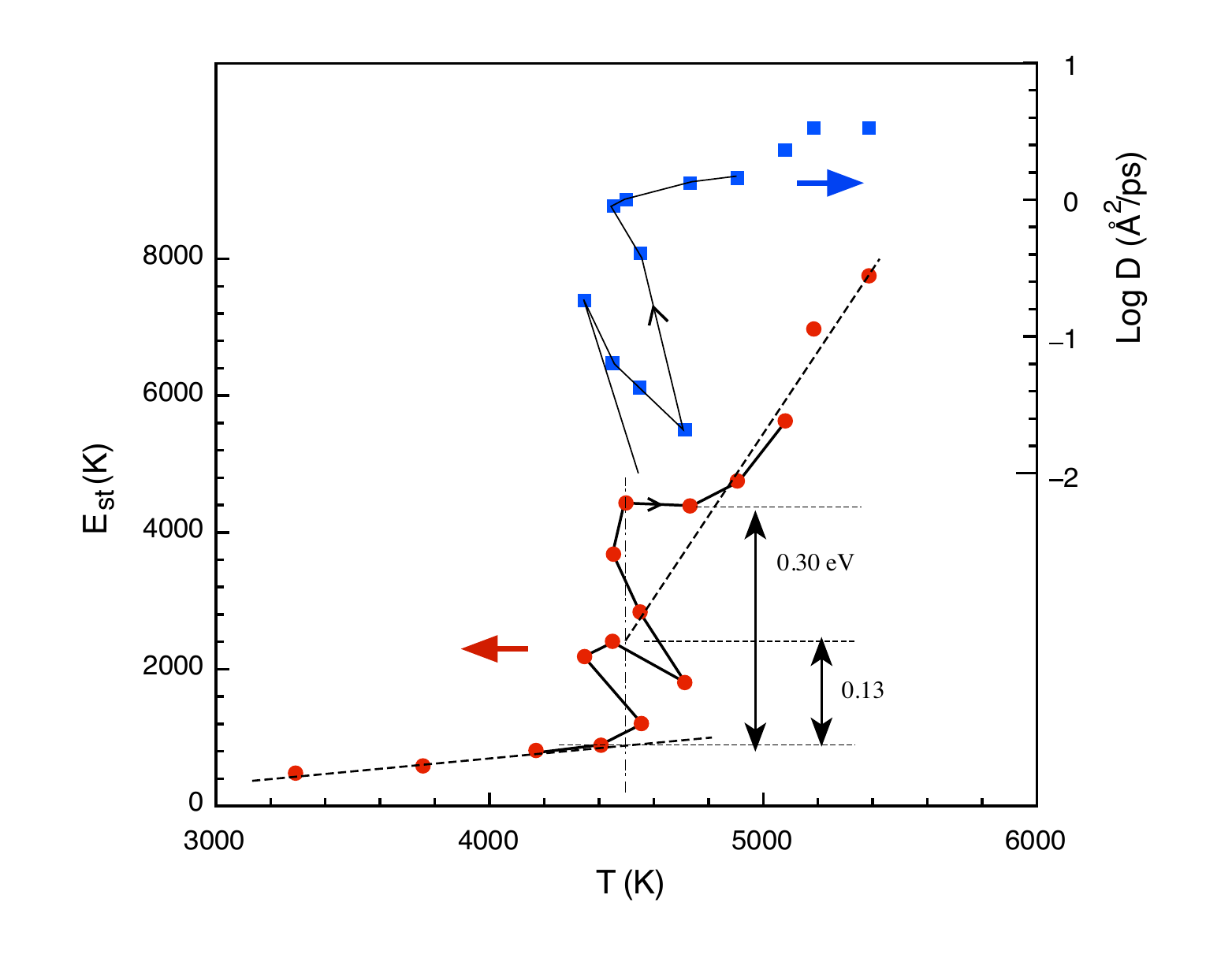} 
  \caption{Structural energy, $E_{\rm st}$, (red circles) and diffusion coefficient, $D$, (blue squares) in the heating process starting from $\alpha$-quartz crystal. $E_{\rm st}$ is plotted in the unit of K per atom. During melting, the sequence of heating is indicated by lines to show oscillatory behavior.
  } \label{fig:heating}
\end{figure}
Figure \ref{fig:heating} shows variations of the diffusion coefficient, $D$, and structural energy, $E_{\rm st}$, as functions of $T$. $D$ is presented in the unit of \AA$^{2}$/ps throughout this paper.
The structural energy, $E_{\rm st}$, is virtually constant with respect to $T$ until $T=4000$ K, indicating that the crystal potential is approximately harmonic. 
Finite values of $D$ appear around $T=4500$ K, indicating the start of melting. In accordance with the behavior of $D$, the structural energy $E_{\rm st}$ abruptly increases. 
The estimated $T_{m}=4500$ K is significantly greater than the experimental value (1600 K). 
The overestimation of melting temperature and similar quantities for silica are commonly observed in MD simulations, regardless of whether empirical \cite{Vollmayr96,Yamahara01,Kuzuu04,Takada04,Geske16,Niu18} or FP potentials \cite{Kim12} are used.
The reasons are discussed from various perspectives.\cite{Chelikowsky01,Yeh04,Sugino95}
Each reason may be correct from the respective perspectives. We hypothesize that the primary reason for the FP potentials is the spurious energy barrier created by the periodic boundary condition with small supercell sizes, which eliminates long-wavelength phonons.\cite{Shirai22-SH}  Appendix A explains this phenomenon.

Figure \ref{fig:heating} shows large fluctuation occurs around $T_{m}$. A close inspection shows that the significance of this fluctuation is more than just random variation. In the figure, the sequence of runs near $T_{m}$ is indicated by connecting points using lines. In this region, the diffusion coefficient, $D$, decreases as $T$ is slightly increased, and vice versa. This is the opposite of the normal behavior---an increase in $T$ increases $D$, implying instability. MD simulations are performed in finite-size cells. This causes a finite width in the energy distribution by $\Delta T/T = 1/\sqrt{N}$. When the structure is changed, the energy distribution between the vibrational and translational motions becomes unbalanced. 
At temperatures slightly above $T_{m}$, atoms with kinetic energy higher than the average kinetic energy, $\langle E_{K} \rangle$, begin to convert their motion to diffusing motion, decreasing $\langle E_{K} \rangle$ and increasing diffusion.
Conversely, at temperatures slightly below $T_{m}$, the low-energy part of atoms, which is absent when $\Delta T=0$, is populated by absorbing energy from diffusing motion. Thus, an oscillatory behavior appears around $T_{m}$.

The step $\Delta E_{\rm st}$ in the structural energy at $T_{m}$ corresponds to the latent heat, $H_{m}$. The $H_{m}$ value is blurred by the large fluctuation in $T$, ranging from 0.13 to 0.30 eV/atom. Even the calculated lower bound of $H_{m}$ is significantly higher than the experimental value, 9.4 kJ/mol (0.03 eV/atom) of $\alpha$-quartz.\cite{Mysen-Richet05} This overestimation of the latent heat is likely to have the same origin as the overestimation in $T_{m}$. The spurious energy barrier created by periodic boundary conditions increases $H_{m}$ and $T_{m}$.

\begin{figure}[htbp]
    \centering
    \includegraphics[width=100mm, bb=0 0 420 470]{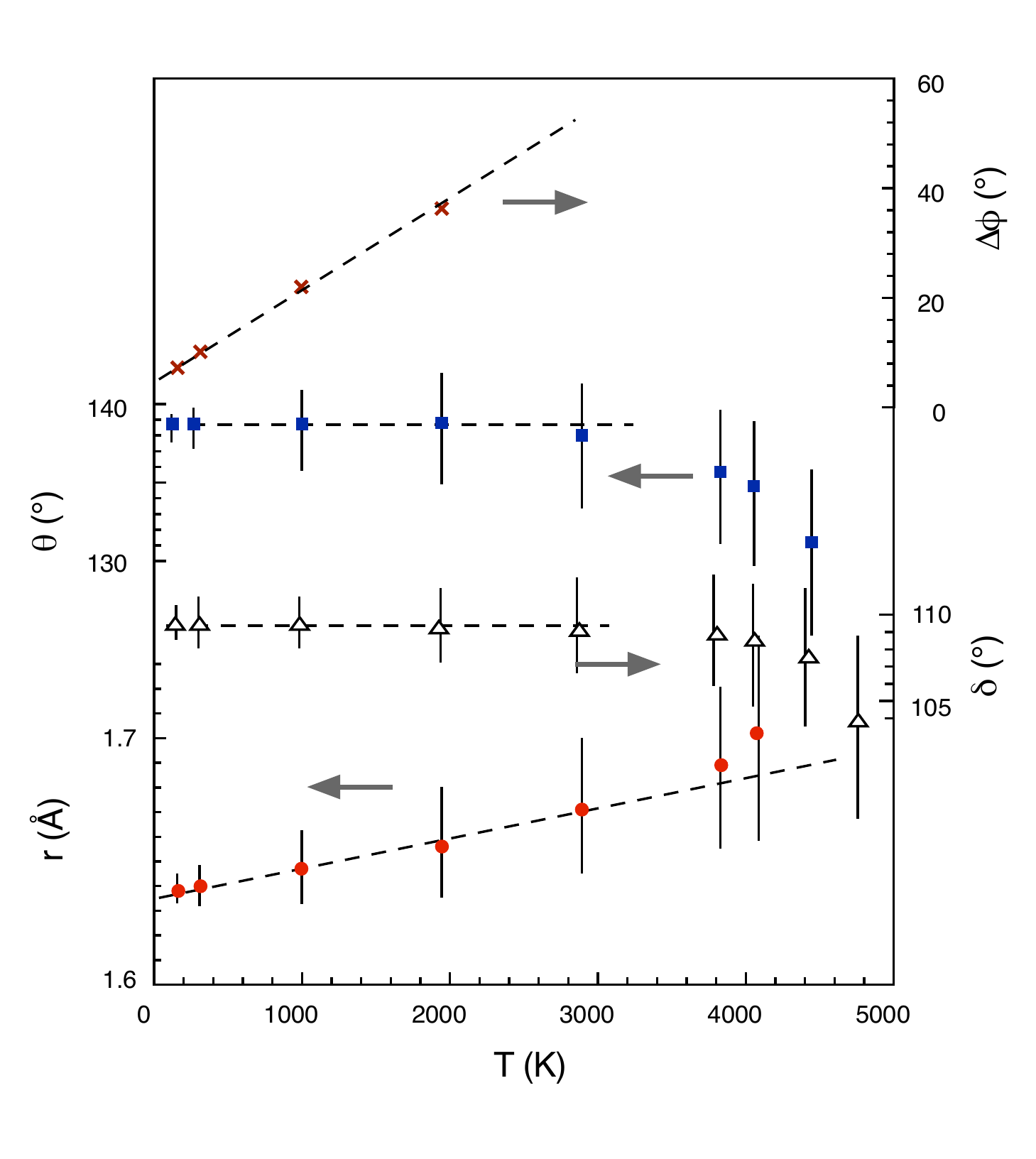} 
  \caption{Structural parameters in the heating process from $\alpha$-quartz: $r$, bond length Si--O; $\theta$, angle Si--O--Si; $\phi$, rotation of O atom around the Si--Si axis. The high-density cell is used to meet the density of $\alpha$-quartz. The mean square variations are represented by bars. Angle variations are reduced by half to make plotting easier.
  } \label{fig:struct-paras}
\end{figure}

The value of $H_{m} =0.03$ eV/atom of silica glass is very small among covalent crystals; for example, silicon crystal has $H_{m} =0.5$ eV/atom despite a similar value of $T_{m}$. This implies that the melting of quartz is due to bond switching, rather than bond breaking.\cite{Mysen-Richet05} This probably explains why silica glass has a high viscosity even in the melt. 
The energy barrier of glass transition, $E_{b}$, is expected to be of the same order as the latent heat, $E_{b} \sim H_{m} \sim RT_{m}$. 
The reality is rather different however; the activation energy, $Q_{a}^{\ast}$, of silica glass is reported to be greater than 5 eV.\cite{Richet84a,Toplis98,Hunt96,Ojovan08} 
This large discrepancy between expectation and experiment has long been known as a big problem in glass physics. This magnitude of $Q_{a}^{\ast}$ is comparable to the energy of covalent bonds. Thus, it is occasionally claimed that the activation energy for the glass transition of silica glass originates from bond breaking (for example, Ref.~\onlinecite{McMillan94}). This interpretation is unrealistic based on the foregoing energetics. The bond energy should be compared with the cohesive energy, which is on the order of magnitude of eV.
Therefore, there must be a mechanism through which the apparent activation energy is obtained experimentally with a substantial overestimation. The mechanism was only recently disclosed by the first author of this study.\cite{Shirai21-ActEnergy} Experimentally, the activation energy is obtained by the Arrhenius plot. When the energy barrier varies rapidly as temperature changes, this temperature dependence appears as a magnification factor, $k=T_{g}/\Delta T_{g}$, for the apparent activation energy.

Some structural parameters in Fig.~\ref{fig:struct-paras} are plotted as a function of $T$. For calculating these parameters, another series of heating processes were performed using high-density cells to achieve the most accurate structural parameters feasible. The shown angles are the angle of O--Si--O ($\delta$) in the tetrahedral unit, SiO$_{4}$, the bending angle of Si--O--Si ($\theta$), and the rotation angle of the O atom ($\phi$) about the Si--Si axis, which are the terminal atoms of the Si--O--Si bond. The bond length, $r$, of Si--O increases slightly with an increase in $T$. The rate of increase is $(1/r)dr/dT = 7.3\times 10^{-6}$/K. 
This value is close to the linear thermal expansivity of $\alpha$-quartz, which is about $1.2 \times 10^{-5}$/K on the orientational average (Ref.~\onlinecite{PSJ-Handbook}, p.~164), even though the current value was obtained by constraining the volume constant. This is discussed in Sec.~\ref{sec:Reheating}. The tetrahedral angle, $\delta$, does not change over a wide range of $T$. Therefore, the tetrahedra SiO$_{4}$ expands uniformly as the temperature increases.

At $T<2000$ K, there is virtually no change in the bending angle, $\theta$, even though the distribution of $\theta$ broadens as $T$ increases in both directions of higher and lower angles. More bending is expected due to the increase in bond length under the constraint of fixed $V$. 
However, there are further freedoms in the parameters in order to allow uniform expansion of tetrahedra SiO$_{4}$ while maintaining a constant volume, $V$. The tilt angle, $\psi$, describes how two adjacent corner-shared tetrahedra can change their relative orientation. This tilt angle, $\psi$, plays an important role in the $\alpha$--$\beta$ phase transition of quartz.\cite{Grimm75} This angle can decouple the change in $V$ from the change in $\theta$.
A relationship exists between the tilt angle, $\psi$, and the rotational angle, $\phi$. The change in $\phi$ is remarkable, as seen in Fig.~\ref{fig:struct-paras}. The angle $\phi$ rapidly increases as $T$ increases. The data displayed are the mean-square variation, $\Delta \phi$, which has a large distribution; some are larger than 90$^{\circ}$, implying virtually unrestricted rotation. 
The fact that the rotation of the bending bond Si--O--Si begins at temperatures significantly lower than $T_{m}$ explains why the $\alpha$--$\beta$ phase transition occurs at such a low temperature ($T=846$ K).

%%%%%%%%%%%%%%%%%%%%%%%%%%%%%%%%%%%%%%%%%%%%%
\subsection{Glass transition from liquid}
\label{sec:cooling}
\subsubsection{Structural energy}
The samples of liquid silica obtained in the preceding section were cooled to obtain the glass.
\begin{figure}[htbp]
    \centering
    \includegraphics[width=120mm, bb=0 0 400 450]{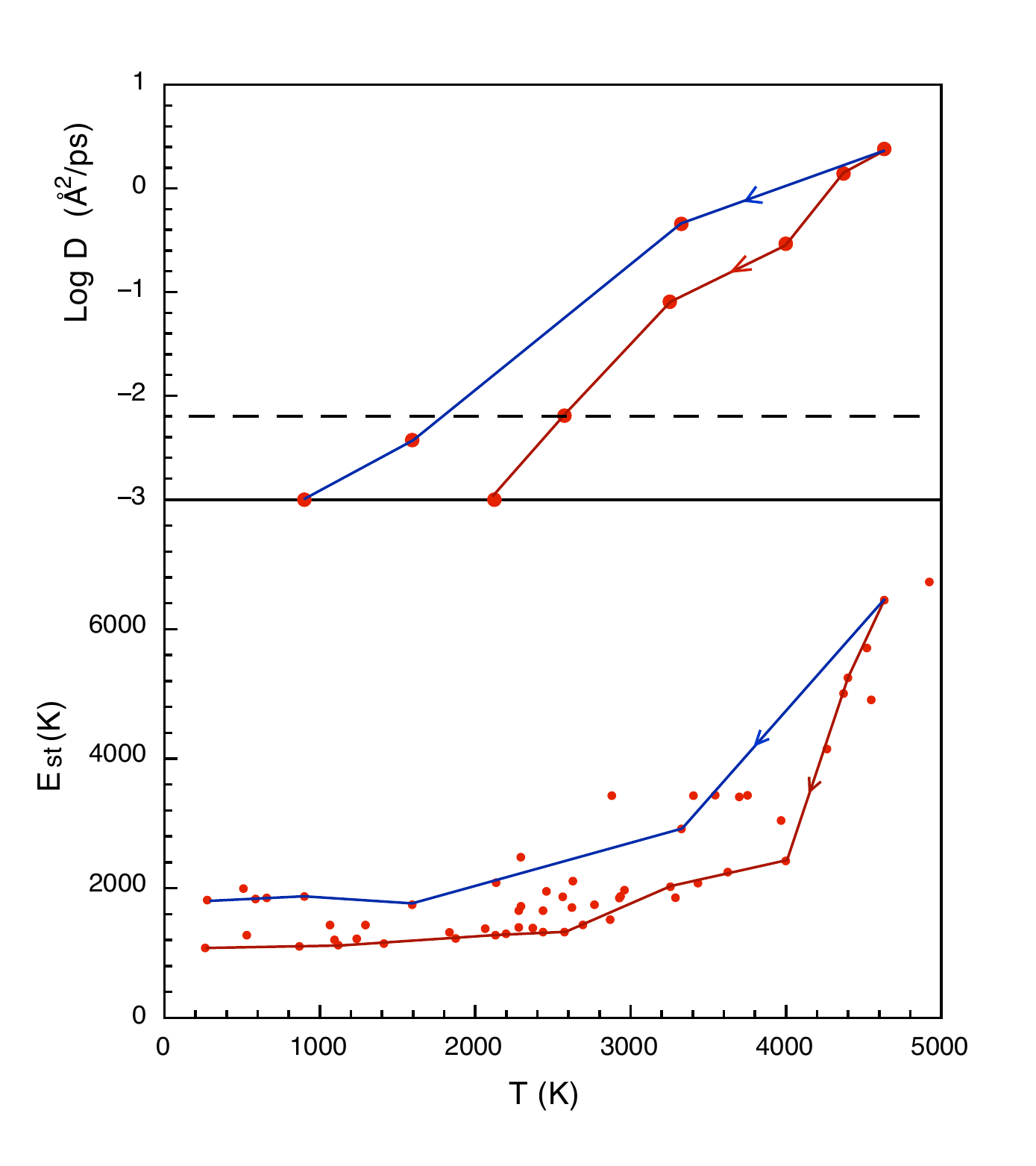} 
  \caption{The structural energy, $E_{\rm st}$, and diffusion coefficient, $D$, in the cooling process: (A) fast cooling indicated by a blue line; (B) slow cooling indicated by a red line. The lines for samples (A) and (B) indicate the sequence of MD runs. For $D$, data of only these two samples are plotted for clarity. The horizontal dashed line indicates the resolution limit for $D$. 
  Data points of $\log D =-3$ actually indicate $D=0$. The energy origin is taken to the ground-state energy of $\alpha$-quartz.
  } \label{fig:Est-D-cooling}
\end{figure}
Several cooling sequences are examined. 
The results are plotted in terms of $E_{\rm st}$ and $D$ in Fig.~\ref{fig:Est-D-cooling}. 
Two contrasting cooling sequences, among others, are chosen as the representatives, and are indicated using lines: (A) fast cooling---as described in Sec.~\ref{sec:Backgrounds}---indicated using a blue line; (B) slow cooling indicated using a red line. Diffusion coefficients in the range of $\log D <-2$ are too small in the current scale and are unreliable. 
Data points of $\log D=-3$ correspond to $D=0$ within the current accuracy; these are plotted only to indicate where glass transition occurs. 
As the temperature is decreased, both $E_{\rm st}$ and $D$ are decreased, as usual. This implies that there is no unstable region as observed in the melting process.
Three regions are identified in the $E_{\rm st}$--$T$ curves. At the high-$T$ region, e.g., $T>4000$ K for sample (A), the gradient $dE_{\rm st}/dT$ provides a significantly large specific heat value, $C_{\rm st} > 5 R$. The temperature is too high compared with $T_{m}$, making the silica melt to be highly volatile. The constraint of constant $V$ could result in unpredictable results, thus we do not discuss this region. In the middle region, e.g., $2600 < T <4000$ K for sample (A), the gradient, $dE_{\rm st}/dT$, has a moderated value: $C_{\rm st} = 0.62 R$ for sample (A), whereas $C_{\rm st} = 0.68 R$ for (B). At the low-$T$ region, e.g., $T<2500$ K for sample (A), the structural energy, $E_{\rm st}$, becomes almost temperature-independent, implying a solid state. 
% However, for sample (B), the average value, $C_{\rm st} = 0.12 R$, is obtained at $T<2000$ K. 
The diffusion coefficient, $D$, vanishes at the boundary between the last two regions. Therefore, the boundary must correspond to the glass transition. For sample (A), $T_{g} = 2600$ K, whereas for sample (B), $T_{g} = 1600$ K. 
After determining $T_{g}$ in this way, the contribution of structural energy to the specific jump was found to be in the range of $\Delta C_{\rm st} = 0.50 R$ to $0.68 R$, depending on the cooling rate. Regarding the transition width, $\Delta T_{g}$, our simulations have such large fluctuation that a clear $\Delta T_{g}$ could not be identified. 

The structural energy, $E_{\rm st}$, decreases as the cooling rate decreases, implying increased stability. This is to be expected because slow cooling makes achieving the energy minimum easier. The difference in $E_{\rm st}$ between samples (A) and (B) is about 800 K (0.07 eV/atom). The rate of cooling in the current MD simulations causes the energy difference to this extent; this may be an overestimation because the cooling rate in the calculation is outside the range accessible in experiment. % of what can be done experimentally. 
The energy difference, $\Delta E_{gc}$, between the crystal and glass sample (B), is 1100 K (0.094 eV/atom), which is three times larger than the experimental value, 9.14 kJ/mol (0.032 eV/atom), reported by Richet {\it et al}.\cite{Richet82} Notably, Ray previously reported that $\Delta E_{gc}=29$ kJ/mol after accounting for the effect of sample grinding.\cite{Ray22} It is unclear why this significant difference in $\Delta E_{gc}$ occurs. We currently trust the value provided by Richet {\it et al}, which implies to accept a significant disparity between the calculation and the experiment. 

\begin{figure}[htbp]
    \centering
    \includegraphics[width=140mm, bb=0 0 768 300]{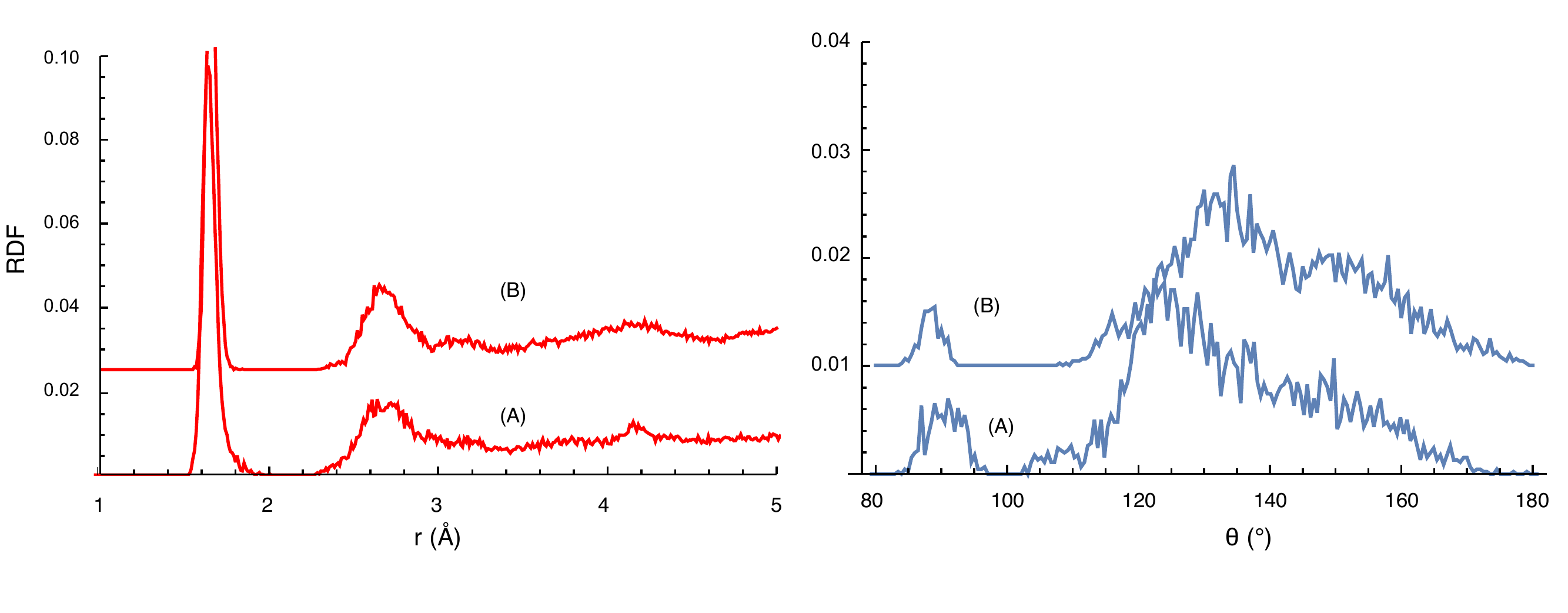} 
  \caption{Comparison of the radial-distribution-function (RDF) and the distribution of angle $\theta$ in samples (A) and (B). The intensity is plotted in arbitrary units, but the scale of the two samples are the same.
  } \label{fig:RDF}
\end{figure}
The aforementioned difference in $E_{\rm st}$ (0.07 eV/atom) between samples (A) and (B) must be attributed to the structural difference. Figure \ref{fig:RDF} shows a comparison of radial-distribution function (RDF) between samples (A) and (B).
There is no discernible difference in RDF between them. However, the distributions of angle $\theta$ are different for different structures. In contrast to the crystal case, $\theta$ has a broad distribution range.
The maximum position of angle $\theta$ is about 10$^{\circ}$ times lower in sample (A) than in sample (B). In addition, a small feature is observed in both samples around $\theta=90^{\circ}$, which was observed in FP calculations by Kim {\it et al.}\cite{Kim12} The low-angle part increases from the sample (B) to sample (A). From the correlation between the energy and angle change, a decrease in $\theta$ (more bending) increases the structural energy, $E_{\rm st}$. In sample (B), structural analysis shows that in a unit cell, one oxygen atom out of 48 has a pair of bond angles with substantial bending around $\theta=90^{\circ}$. 
If this pair of angle defects were removed, $E_{\rm st}$ would be further reduced. 
We attempted to remove this defect by repeating the annealing and cooling process, without success. Perhaps the small size of the cell (72 atoms) is too restrictive to remove this defect.

\begin{figure}[htbp]
    \centering
    \includegraphics[width=100mm, bb=0 0 380 280]{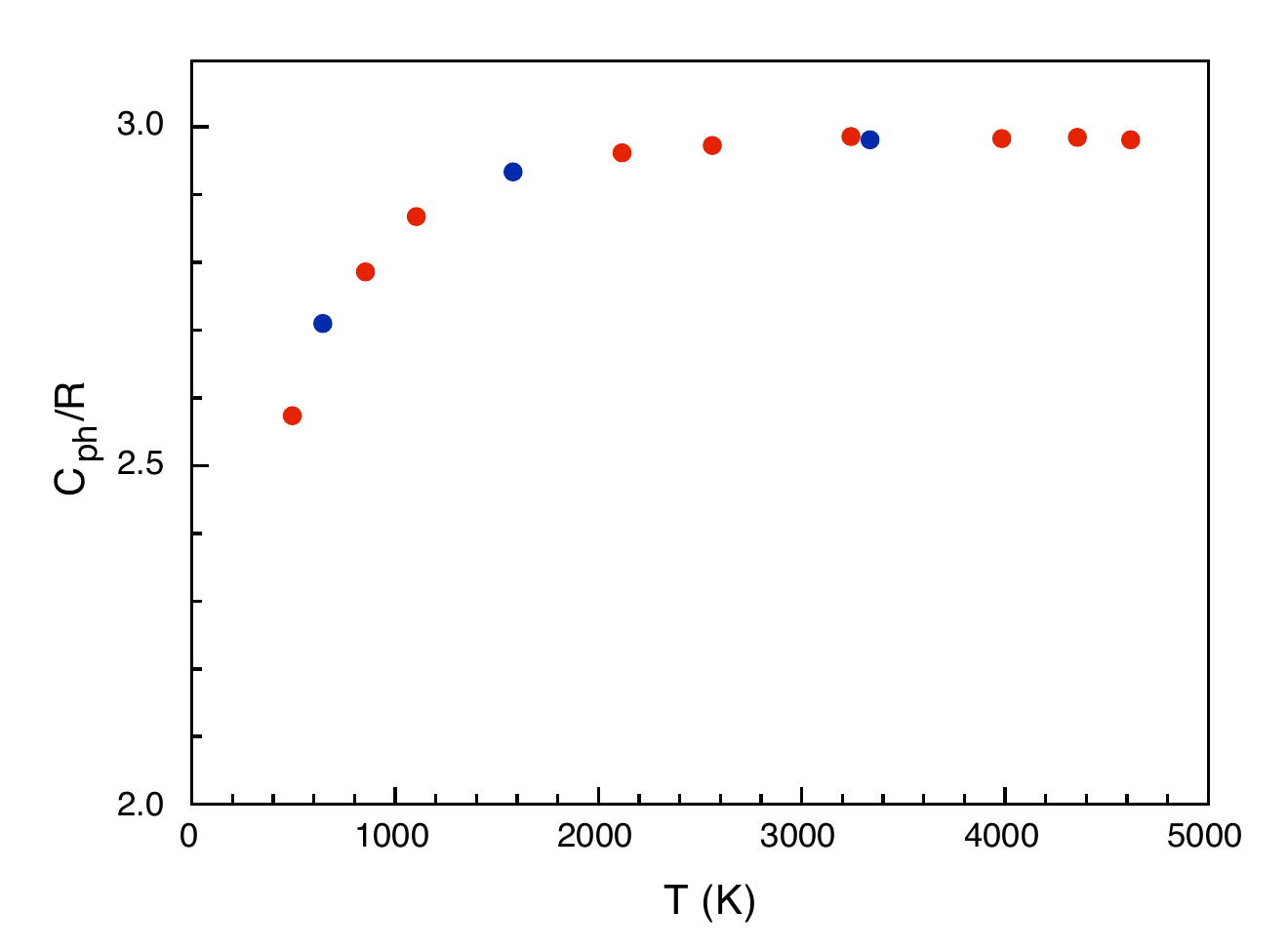} 
  \caption{Phonon contribution to specific heat $C_{\rm ph}$ of silica glass. Blue: fast cooling sample (A), red: slow cooling sample (B).
  } \label{fig:Cp-T-calc}
\end{figure}

\subsubsection{Phonon contribution}
\label{sec:phonons}
Next, let us investigate the phonon contribution to the specific heat.
Figure \ref{fig:Cp-T-calc} shows the specific heat of the phonon part, $C_{\rm ph}$. The figure shows that $C_{\rm ph}$ is close to the classical limit, $3R$, at high temperatures $T>2000$ K, thus the specific jump cannot be seen. If any is observed, then it is less than the current resolution limit of about $0.01R$.
\begin{figure}[htbp]
    \centering
     \includegraphics[width=90mm, bb=0 0 380 520]{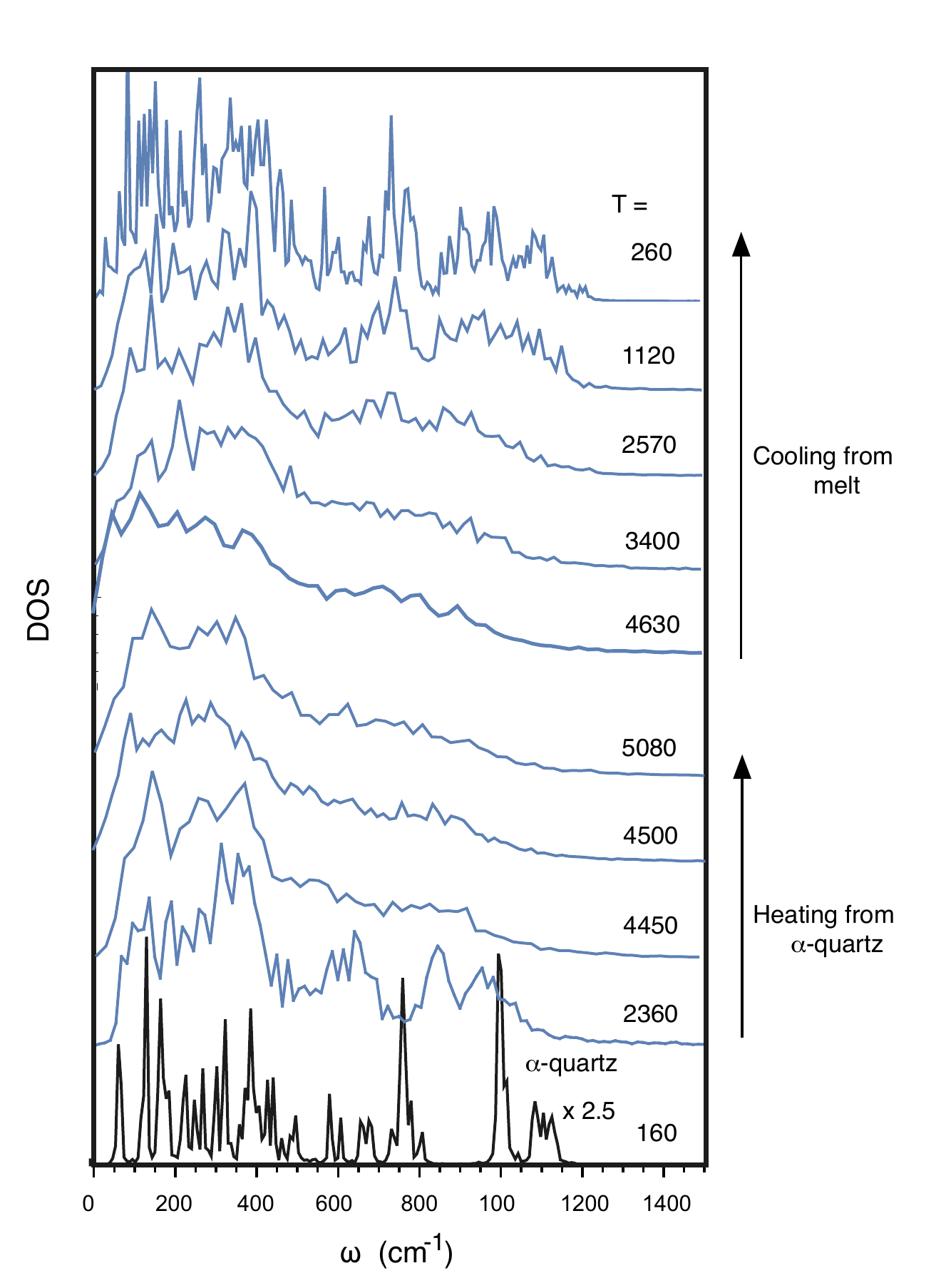} 
  \caption{
  Evolution of phonon spectra in the melting process of crystal $\alpha$-quartz and in the quenching process from silica melt. The intensity of crystal $\alpha$-quartz at $T=160$ K is reduced by 2.5 for easy visibility.
  } \label{fig:phonspectra}
\end{figure}

This negligibly small value of $C_{\rm ph}$ can be understood by examining the evolution of the phonon spectra. Figure \ref{fig:phonspectra} shows the evolution of phonon spectra throughout the melting and cooling processes. Readers may observe shifts in the calculated spectra to lower frequencies if they compare the experimental spectra of crystal $\alpha$-quartz \cite{Shapiro72} and silica glass \cite{Galeener83}. However, this is due to the TO--LO splitting for polar modes, which is not considered in the calculations. 
The phonon spectrum of crystal $\alpha$-quartz comprises multiple bands that are roughly divided into three groups: a high-frequency band at about 1100 cm$^{-1}$, a middle-frequency band at about 800 cm$^{-1}$, and a broad low-frequency band from 200 to 450 cm$^{-1}$. The fine structure of the crystal broadens as temperature increases, whereas these main features remain. We also observe traces of the main features even for the liquid. The phonon spectra shift slightly to the low-frequency side, as $\alpha$-quartz is heated up to melt. On cooling the liquid to glass, the phonon spectra take on features similar to those of crystal $\alpha$-quartz. The phonon spectrum of silica glass is well characterized (see paper by McMillan {\it et al.}~for an overview of current knowledge \cite{McMillan94}).
Researchers noticed an unusual behavior at the 440-cm$^{-1}$ band: the frequency increases as $T$ increases.\cite{Geissberger83,McMillan94} This is interpreted to be due to the narrowing of the bending angle of the Si--O--Si bond.\cite{Sen77,Galeener79} Unfortunately, this increase is not resolved in our simulation due to substantial fluctuations in phonon spectra.
In any case, the details of phonon spectra are irrelevant for specific heat.
Because all the phonons are already thermally activated at high temperatures around $T_{g}$, a slight change in $T$ around $T_{g}$ does not affect the temperature dependence of $C_{\rm ph}(T)$, thus there is no reason to expect any jump $\Delta C_{\rm ph}$.

\subsubsection{Effect of thermal expansion}
\label{sec:effectTE}
The third contribution to the total specific heat is from thermal expansion, $C_{\rm te}$. The contribution of thermal expansion is expressed by Eq.~(\ref{eq:Cthermal-exp}). 
\begin{table}
\begin{tabular}{l  rr c l}
  \hline \hline
 Property & \multicolumn{3}{c}{Value}  &  Ref. \\ \cline{2-4}
   & ($g$) & ($l$)  & unit &  \\
   \hline
  $V$ & 27.27 & \ 27.23  &  cm$^{3}$/mol  & \cite{Mysen-Richet05} \\ 
  $C_{p}$ & 73.40 & 81.37 & J/mol$\cdot$deg & \cite{Richet82} \\ 
  $\alpha$ &  1.8 & $+ 2 \%$  & $\times 10^{-6}$/K & \cite{Bruckner71,Oishi69,Kuhn09} \\ 
  $\kappa$ &  22 & $- 2 \%$ & TPa$^{-1}$ & \cite{Polian02,Guerette12} \\ 
    \hline  
\end{tabular}
\caption{Thermodynamic data of silica glass used in this study; $V$: molar volume, $C_{p}$: specific heat, $\alpha$: volume thermal expansivity, $\kappa$: isothermal compressibility. In column ($l$), $+x\%$ means that $A^{(l)}=A^{(g)}(1+0.01 x)$ for any quantity $A$.}
\label{tab:Properties-silica}
\end{table}
Experimental data from the literature was used to evaluate this formula. The used values are tabulated in Table \ref{tab:Properties-silica}. Because we are interested in the change in $C_{\rm te}$ around $T_{g}$, the values must be those just below and above $T_{g}$. However, measurements near $T_{g}$ are rather scarce. Determining $C_{\rm te}$ for silica glass suffers several difficulties.
First, data on silica glass vary among authors. The current authors arrived at the cited values after a comparison of various data. 
Second, the changes in thermodynamic properties near $T_{g}$ are extremely small to be determined accurately. Third, because the width of the glass transition, $\Delta T_{g}$, for silica glass is not well established, the changes in properties largely depend on the $\Delta T_{g}$ value used.

The reported data for volume expansivity, $\alpha$, are rather well converged in a range of 1.3--2.1$\times 10^{-6}$/K near room temperature.\cite{Bruckner71,Oishi69,Kuhn09} However, there are few measurements of $T$ dependence of $\alpha$ around $T_{g}$ and the reported data have substantial fluctuations, making it difficult to evaluate $\Delta \alpha$.\cite{Kuhn09} Using $\Delta T_{g}=100$ K (see the data by Sudo in Fig.~\ref{fig:CTexperiment}) and the expansivity data reported by Kuhn and Schadrack,\cite{Kuhn09} an amount of $\Delta \alpha/\alpha = 2 \%$ was estimated.
The compressibility of silica glass is the most confusing quantity. The issue is the large disparity in $\kappa$ between the static and dynamic measurement---the static value is about four times larger than the dynamic value.\cite{Bucaro76,Kress89} Mysen and Rechet reviewed previous data with appropriate interpretation.\cite{Mysen-Richet05} In this study, we used recent values of Brillouin scattering measurement by Polian {\it et al.}\cite{Polian02} and Guerette {\it et al.}\cite{Guerette12} because the compressibility was measured over a wide range of $T$ covering near $T_{g}$. The compressibility, $\kappa$, increases unusually as $T$ is increased. However, the increase is too small in the transition region to accurately determine the jump $\Delta \kappa$. The jump $\Delta \kappa/\kappa = -2 \%$ has been estimated at most by assuming $\Delta T_{g}=100$ K.

From these values listed in Table I, the thermal expansion part of specific heat, $C_{\rm te}$, is determined as $C_{\rm te}^{(g)} = 2.3 \times 10 ^{-4}R$ and $C_{\rm te}^{(l)} = 2.5 \times 10 ^{-4}R$ for glass and liquid, respectively, with a very small difference, $\Delta C_{\rm te} = 2.\times 10^{-5}R$. Therefore, the contribution of thermal expansion to $\Delta C_{p}$ is negligible. 

\subsubsection{Total specific heat}
\label{sec:totalC}
By analyzing the three components, we conclude that the contribution from structural energy, $E_{\rm st}$, is responsible for the specific-heat jump of silica glass. 
By considering similar conclusions in other studies on different glasses,\cite{Smith17,Han20,Shirai22-SH} although the degree of contribution of $\Delta C_{\rm st}$ is different, it is certain that the property that $\Delta_{p}$ is predominated by $\Delta_{\rm st}$ is a general property of glasses.
The value of $\Delta C_{p}$ of silica glass varies between $0.50 R$ and $0.68 R$, depending on the cooling rate. This is an overestimation of the experimental value of $0.32 R$, which is probably due to the spurious energy barrier of the small-size supercell.
The jump of $0.32 R$ of silica glass is small compared to the jump in $\Delta C_{p}$ of fragile glasses, which is of the order of $R$. This does not mean, however, that the change in the structural energy, $E_{\rm st}$, of silica glass is insignificant. 
Specific heat represents the fluctuation in the microscopic energy, as similar to entropy $S$. Entropy becomes insensitive to the change in $T$ at high temperatures, as is seen in the relation $\Delta S=\Delta U/T$. Because the $T_{g}$ of silica glass is as high as 1480 K, the small $\Delta C_{p}$ is a consequence of the insensitivity of entropy at high temperatures.

Finally, although we are not much concerned about it, let us verify the specific heat of liquid. The current simulations show that $E_{\rm st}$ for the liquid is almost linear, indicating constant $C_{\rm st}^{(l)}$; the calculated value is $C_{\rm st}^{(l)}=0.73 R$. By considering that the phonon contribution has already reached the classical limit, $3R$, and that thermal expansion is negligible, we conclude that the total specific heat of the liquid is constant, with $C_{p}^{(l)}=3.73R$. Richet {\it et al.}~reported a constant value, $C_{p}^{(l)} = 3.26 R$, in a temperature range of $1480 < T < 2000$ K.\cite{Richet82} Despite this overestimation, our result is consistent with theirs in terms of $C_{p}^{(l)}$ being a constant. 
However, we cannot assert that the calculation is correct simply because it agrees with Richet {\it et al.}'s experiment. Our simulations overlook the component of low-frequency phonons, whose frequency is less than 50 cm$^{-1}$, due to the limited size of the supercell. These low-frequency phonons can be converted to purely translational motions as $T$ increases. Thus, contrary to being constant, a reduction in specific heat from $3R$ to $1.5R$ is expected if the temperature range is sufficiently large.\cite{Wallace02} 
Alternatively, we do not have enough reason to rule out the experimental finding of an increase in $C_{p}^{(l)}$ with increasing $T$. 
Currently, we can only say that thermodynamic properties of even silica liquids have hysteresis in the range of $1500 < T <2000$ K, as suggested by researchers of glass industry.

%%%%%%%%%%%%%%%%%%%%%%%%%%%%%%%%%%%%%%%%%%%
\subsection{Reheating of glass}
\label{sec:Reheating}
\begin{figure}[htbp]
    \centering
    \includegraphics[width=95 mm, bb=0 0 400 420]{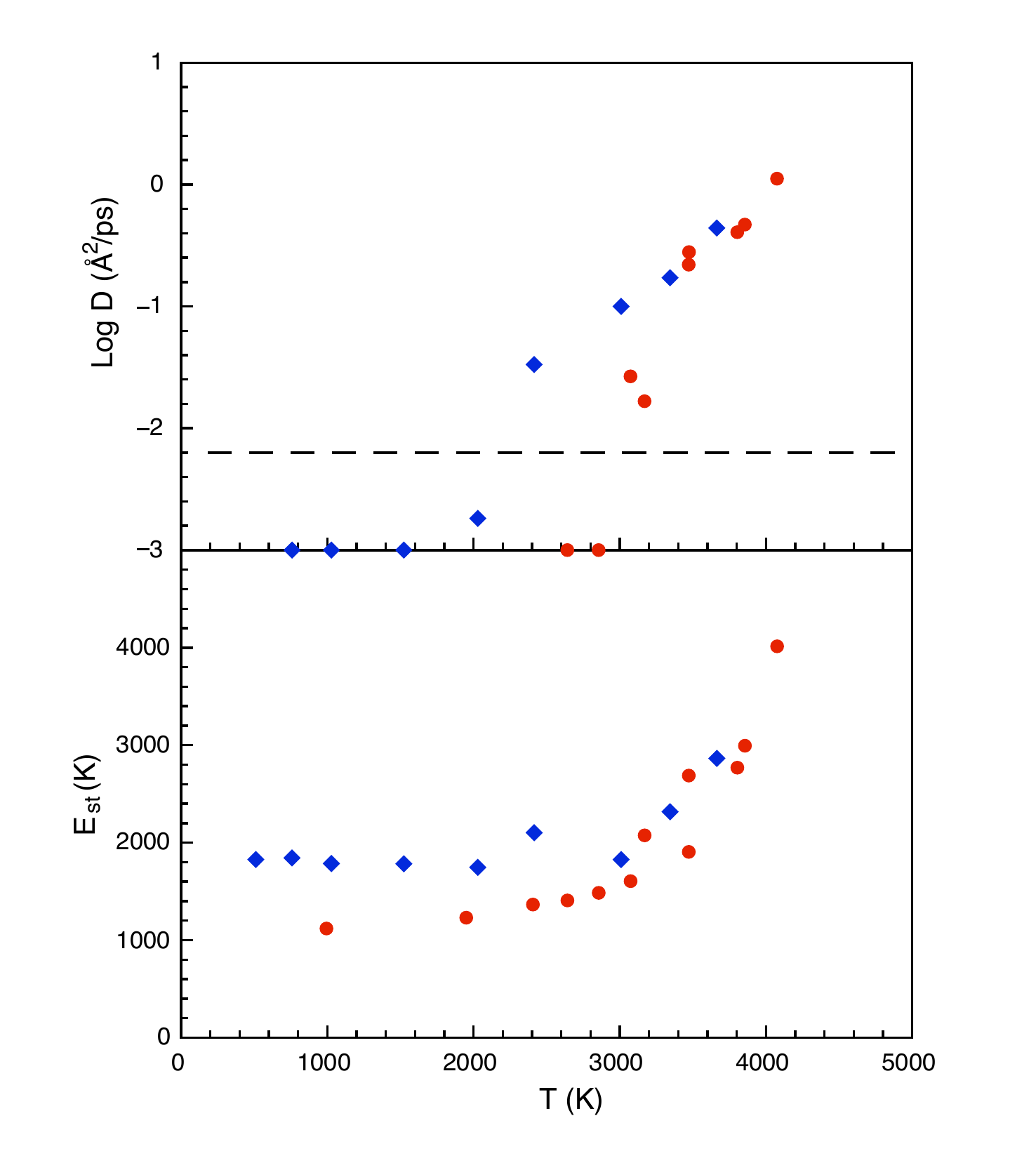} 
  \caption{Diffusion coefficient, $D$, and structural energy, $E_{\rm st}$, of silica glass in reheating process. Two samples, A (blue) and B (red), are compared.
  } \label{fig:Est-reheat}
\end{figure}
The obtained glasses (A) and (B) were heated again from low temperatures to examine the dependence on processes. Figure \ref{fig:Est-reheat} shows variations in the diffusion coefficient, $D$, and structural energy, $E_{\rm st}$, when the glass is heated. The figure shows that the glass transition temperature is higher in this instance than in the cooling scenario: $T_{g}$ is about 1600 K for sample (A) and 3200 K for sample (B). 
This increase is expected because the barrier height is higher when measured from low temperatures. Hence, we trust the $T_{g}$ values obtained in the cooling process. 
The calculated hysteresis between the cooling and reheating processes is, however, different from the experimentally observed one because the latter is caused by the inhomogeneity of samples, whereas the former is due to the artificial boundary condition.

\begin{figure}[htbp]
    \centering
    \includegraphics[width=100mm, bb=0 0 420 470]{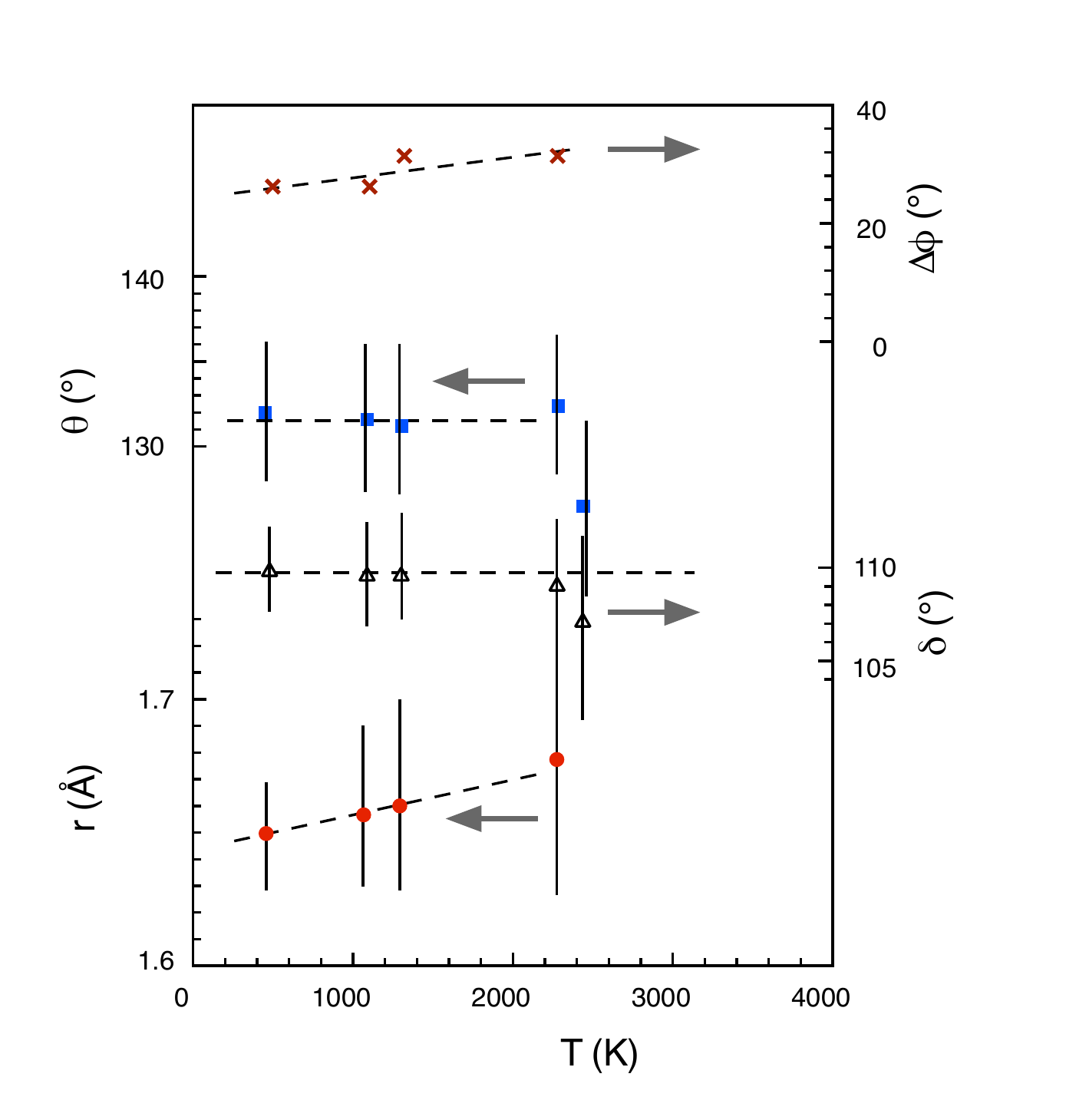} 
  \caption{Structural parameters of silica glass in the reheating process: $r$, bond length Si--O; $\theta$, angle Si--O--Si; $\phi$, rotation of O atom around the Si--Si axis. The low-density cell is used to meet the density of silica glass. The notations are the same as in Fig.~\ref{fig:struct-paras}.
  } \label{fig:struct-paras-reheat}
\end{figure}
Figure \ref{fig:struct-paras-reheat} shows the temperature dependence of the structural parameters. 
The change in bond length against $T$ is $(1/r)dr/dT = 7.9\times 10^{-6}$/K, which is close to that of $\alpha$-quartz obtained in the heating process. This implies that the change in bond length is irrelevant to the small thermal expansivity of silica glass.
The constancy in angles $\delta$ and $\theta$ with respect to $T$ is also observed in crystal $\alpha$-quartz. This constancy of $\theta$ is different from that obtained by model potentials,\cite{Yamahara01, Takada04, Kuzuu04} where $\theta$ decreases as $T$ is increased.
Although the mean value of $\theta$ is constant with respect to $T$, a notable difference is its large mean-square variation. Figure \ref{fig:RDF} shows that the bending angle, $\theta$, of the Si--O--Si bond has a wide range from 80 to 180$^{\circ}$. Similarly, the mean-square variation, $\Delta \phi$, is also large even at room temperature.
As mentioned in Sec.~\ref{sec:phonons}, a decrease in $\theta$ from 140$^{\circ}$ increases the frequency of the bond stretching mode perpendicular to the Si--Si axis, whereas an increase from 140$^{\circ}$ results in an opposite outcome.\cite{Sen77,Galeener79} 
This effect is observed in the shift in the 440-cm$^{-1}$ band of the Raman spectra of the crystal $\alpha$-quartz. \cite{McMillan94}
Both directions of expansion and contraction of the unit cell, which counterbalance each other, are caused by two parts of higher and lower angle $\theta$ relaxing thermal strains. This could explain why silica glass has a far lower thermal expansivity than crystal $\alpha$-quartz, despite their structural similarities. However, we do not go further on this topic because there are already numerous studies on it.

%%%%%%%%%%%%%%%%%%%%%%%%%%%%%%%%%%%%%%%%%%
\section{Interpretation of the Prigogine-Defay relation}
\label{sec:discussion}

The preceding results provide insight on the Prigogine-Defay relationship. The following relationship holds for glasses:
\begin{equation}
{\it \Pi} = \frac{\Delta C_{p} \Delta \kappa}{TV (\Delta \alpha)^{2}} \geqq 1,
\label{eq:PDrelation}
\end{equation}
where ${\it \Pi}$ is the PD ratio. The formal confirmation of this relationship has been extensively studied \cite{Davies53, Goldstein73,DiMarzio74,Gupta76,Goldstein75,Lesikar80}, but the subtlety of this relationship is still up for debate \cite{Nieuwenhuizen97,Tropin12, Schmelzer12, Garden13}.
Apart from these subtle problems, it is a general understanding that equality in Eq.~(\ref{eq:PDrelation}) only holds when there is just one order parameter; otherwise, inequality holds. The overwhelming experimental data on glasses indicate that ${\it \Pi} > 1$, and in most cases, $2 < {\it \Pi} < 8$, implying the presence of several (or more) order parameters. The abstract nature of the order parameters makes it difficult to derive physical meaning from them. The fact that there appears to be no general trend in ${\it \Pi}$ among various glasses nor correlation to other qualities such as fragility, adds further difficulties. 

In this regard, the novel perspective of state variables of solids proposed by Shirai \cite{Shirai20-GlassState,StateVariable} has significant advantages from which the inequality of (\ref{eq:PDrelation}) is reasonably understood. In the following, we present an exposition of the PD ratio considering this perspective.
Shirai demonstrated that the thermodynamic state variables (thermodynamic coordinates) of a solid are equilibrium positions, $\{ \bar{\bf R}_{j} \}$, of the atoms that comprise the solid.\cite{Shirai20-GlassState} In fact, the set of $\{ \bar{\bf R}_{j} \}$, in addition to $T$ and $V$, forms the arguments of the function $U$ in Eq.~(\ref{eq:internal-energy-3}). This conclusion was deduced from the basic requirements for state variables: first, a state variable must have a definite value in equilibrium; second, an equilibrium state must be uniquely specified by a set of state variables.\cite{StateVariable}
The first requirement differentiates solid states from gas states; the atoms for solids maintain their own and unique positions in any of their equilibrium states, whereas atom positions for gases are indeterminate on time averaging and therefore cannot be state variables. The second requirement guarantees that thermodynamic states are independent of the history in which the current state was obtained. Considering two samples (A) and (B) in Sec.~\ref{sec:cooling}, if the state variables of glass were only $T$ and $V$, we would have to argue that the obtained glass is a nonequilibrium state because there is no function, $U(T, V)$, that describes the difference between these two samples. If the state were in nonequilibrium, we could extract work from it without harming the environment. This conflicts with the second law of thermodynamics. Accordingly, the glass state must be in an equilibrium state. Affirmatively, we can use $U=U(T,V, \{ \bar{\bf R}_{j} \})$ to describe the difference in the thermodynamic properties of samples (A) and (B), irrespective of the history.

Order parameters are nonvanishing and definite values in equilibrium, thus they satisfy the first requirement for state variables. Order parameters are parameters that characterize the structure of a state of matter.\cite{Landau-SP} The structure is best described by a complete set of atom positions $\{ \bar{\bf R}_{j} \}$. Therefore, any property that is uniquely determined by $\{ \bar{\bf R}_{j} \}$ or the entire set $\{ \bar{\bf R}_{j} \}$ can be used for order parameters. We will select the latter choice as our order parameters. This choice satisfies the second requirement for state variables. Thus, the order parameters become equivalent to the state variables of a solid.
This implies that the number of order parameters for glass equals the number of atoms in the glass.
From this viewpoint, inequality is a natural consequence of order parameters in glasses.

We can go further. Silica glass is a special case in that its ${\it \Pi}$ value is extremely high, ranging from $10^{3}$ to $10^{5}$.\cite{Nemilov-VitreousState} The wide range of the ${\it \Pi}$ value is due to the extremely small thermal expansivity of silica glass---the smallest of all materials on the earth. It is difficult to resolve minor changes in the volume of silica glass contained in an apparatus composed of other materials with $\alpha$'s larger than that of silica glass.
By combining Eq.~(\ref{eq:Cthermal-exp}) with Eq.~(\ref{eq:PDrelation}), we obtain the following:
\begin{equation}
{\it \Pi} = \frac{ \Delta C_{p}/C_{\rm te} } { 
    \left( \Delta \alpha/ \alpha \right)^{2} /(\Delta \kappa/\kappa) } .
\label{eq:PDratio1}
\end{equation}
A simple formula for thermal expansivity, $\alpha$, is $\alpha = \gamma C \kappa/V$, where $\gamma$ is the Gr\"{u}neisen parameter, which indicates the magnitude of anharmonicity of atom potentials; for example, Eq.~(25.19) of Ref.~\onlinecite{AshcroftMermin}. In this formula, $\gamma C$ is an averaged value across all phonon modes, assuming that all phonon modes contribute equally to thermal expansion.
However, we must consider the assumption in this case.
Because the thermal expansion is an isotropic response to heat injection, modes that do not violate this isotropy do not contribute to thermal expansivity. Many angle-bending modes are examples of this. An illustrative example is the librational mode of $\alpha$-boron.\cite{Shirai98} Despite the angle-bending forces in $\alpha$-boron having large anharmonicity, the frequency of the librational mode does not have pressure dependency, which is usually considered evidence for small anharmonicity.
Therefore, in the formula for $\alpha$, specific heat, $C$, should be replaced with $C_{\rm is}(V)$, which is contributed only by those phonons whose frequencies depend only on the isotropic volume change.
\begin{equation}
\alpha=  \frac{ \overline{\gamma_{\rm is}(V) C_{\rm is}(V)}}{V} \kappa,
\label{eq:TEform}
\end{equation}
where $\gamma_{\rm is}(V)$ is the corresponding mode-Gr\"{u}neisen parameter.
Notably, the isothermal compressibility, $\kappa$, of silica glass is not so small as is expected from the small thermal expansivity, $\alpha$. This apparent conflict between $\kappa$ and $\alpha$ behavior is due to the unusually small isotropic-Gr\"{u}neisen parameter $\overline{\gamma_{\rm is}}$ of silica glass---it is smaller than usual by one order of magnitude.\cite{Phillips81} The small value of the average Gr\"{u}neisen parameter is due to the counterbalance between positive and negative mode-Gr\"{u}neisen parameters $\{ \gamma_{q} \}$, as described in Sec.~\ref{sec:Reheating}.

When the factor, $\overline{\gamma_{\rm is} C_{\rm is}}/V$, with respect to $T$ is constant, 
the change in $\alpha$ is solely determined by $\kappa$, and their relative changes are the same.
\begin{equation}
\frac{\Delta \alpha}{\alpha} = \frac{\Delta \kappa}{\kappa}.
\label{eq:alpha-kapp}
\end{equation}
This relationship holds for crystals on the same order in anharmonic perturbations; both $\alpha$ and $\kappa$ have linear dependence on $T$ at high temperatures.\cite{Leibfried61}
Similarly, from Eq.~(\ref{eq:Cthermal-exp}), the part of thermal expansion, $C_{\rm te}$, changes as follows:
\begin{equation}
\frac{\Delta C_{\rm te}}{C_{\rm te}} = \frac{\Delta \alpha}{\alpha}.
\label{eq:del_Cte}
\end{equation}
From Eqs.~(\ref{eq:alpha-kapp}) and (\ref{eq:del_Cte}),
Eq.~(\ref{eq:PDratio1}) can be rewritten as
\begin{equation}
{\it \Pi} = \frac{ \Delta C_{p}}{\Delta C_{\rm te}}
    = \frac{ 
    \text{(Change in the total energy)} }{ \text{(Contribution of isotropic volume change)} }.
\label{eq:PDratio3}
\end{equation}
% Alternately, $1/{\it \Pi}$ denotes the contribution of thermal expansion to the change in the total energy at the glass transition. 
Because ${\it \Pi} > 1$ (mostly $>2$), the glass transition occurs mainly due to a change in the internal structure that determines $E_{\rm st}$: the contributions of thermal expansion and phonons are negligible.
Section \ref{sec:results} shows that the jump $\Delta C_{p}$ for silica glass is almost entirely determined by $\Delta C_{\rm st}$, leading to ${\it \Pi} \gg 1$. As noted in Sec.~\ref{sec:totalC}, specific heat as well as entropy reflect energy fluctuations, which scale to $T$. In this sense, ${\it \Pi}$ is better than $C_{\rm st}$ itself for indicating the contribution of structural energy, because the insensitivity to temperature change is eliminated by taking the ratio $\Delta C_{p}/\Delta C_{\rm te}$. As stated in Introduction, the apparent properties of silica seem not to change so much between the glass and liquid states. However, the difference is clearer by looking at ${\it \Pi}$, indicating the significant change in the internal structure.

Filipovich had already attributed $\Delta C_{p}$ of silica glass to the contribution of structural parameters more than 3 decades ago.\cite{Filipovich89} In addition to the volume change, he modeled the $\Delta C_{p}$ of silica glass by changing the distribution of angle Si--O--Si. These two types of changes were described by two independent force constants. From this model, he showed that $\Delta C_{p}$ and $\Delta \alpha$ change independently, resulting in ${\it \Pi}>1$. His model is essentially correct, with reducing the entire set of state variables, $\{ \bar{\bf R}_{j} \}$, to only two variables.

Contrarily, if $\Delta C_{p}$ is caused by only the contribution of thermal expansion, $\Delta C_{\rm te}$, ${\it \Pi} = 1$ is established. This occurs when the changes in thermodynamic properties of the glass are described by elastic models with single parameters, such as the volume change, $\Delta V$. A simple model, such as the free-volume model, is often used in the glass literature. The free-volume model describes the energy change by the effective volume of a ``hole''---the disordered structure is represented by holes. In this case, the equality ${\it \Pi} = 1$ is necessarily established.
Starting with a simple model is a good idea. However, for future development, we cannot rely on the most basic model. 

% Lastly, requiring Eq.~(\ref{eq:alpha-kapp}) is equivalent to imposing Goldstein's condition for equality in Eq.~(\ref{eq:PDrelation}) to hold.\cite{Goldstein73} In the $T$--$P$--$V$ diagram presented in his study (Fig.~3 of Ref.~\onlinecite{Goldstein73}), glass has several branches of sheets, $V=V(T,P)$, each of which corresponds to samples with different densities. He showed that, if the state of the glass remains in one branch, the equality, ${\it \Pi}=1$, holds. Staying in one branch entails that the state is defined solely by two state variables, $T$ and $P$, and that no other state variables are present. In real situations, the equation-of-state has many variables, as shown by $V=V(T,P, \{ \bar{\bf R}_{j} \})$. 

%%%%%%%%%%%%%%%%%%%%%%%%%%%%%%%%%%%%%%%%%%
\section{Summary}
The glass transition of silica glass was investigated using FP-MD simulations by calculating the specific heat. The current simulations are severely limited due to the small size of the cells. The periodic boundary condition with small cell sizes yields an erroneous energy barrier for atom motions, resulting in overestimation in melting temperature, glass-transition temperature, and other similar properties. Despite these shortages, this study provides new information on the thermodynamic properties of silica glass.

The glass-transition temperature, $T_{g}$, was obtained in the range of 1600--2600 K, with a jump in the specific heat, $\Delta C_{p} = 0.50$ to $0.68R$. Although these values are higher than the experimental values, the obtained values are within a reasonable range when the aforementioned uncertainties are considered.
According to the state variables, decomposing the internal energy into three components (structural, phonon, and thermal expansion energies) reveals that the jump $\Delta C_{p}$ of silica glass is entirely determined by the structural energy, $E_{\rm st}$. The change in $E_{\rm st}$ is controlled primarily the distribution of the bending angle of Si--O--Si bond. The reason for the small $\Delta C_{p}$ is the high $T_{g}$ of silica glass, which makes the fluctuation in $E_{\rm st}$ insensitive to the change in $T$.
% The viscosity may be affected by the rotation of the O atom around the Si--Si axis.

One outcome of this study is finding a physical interpretation for the Prigogine-Defay ratio, $\it{\Pi}$. It represents the ratio of the total energy change in the glass transition to thermal expansion contribution. The experimental fact that ${\it \Pi}>1$ implies that the glass transition is mostly caused by the change in $E_{\rm st}$, which is controlled by the internal structure. An extreme example is the current case of silica glass: the glass transition occurs without assistance from the change in thermal expansion or phonons. Despite the apparent similarities between the glass and liquid states, the two states of silica glass are very distinct in their internal structures, which is reflected in the large value of $\it{\Pi}$.

\section*{Acknowledgment}
The authors thank N. Kuzuu (Univ.~Fukui) and P. Richet (Inst.~Phys.~Globe de Paris) for valuable discussion on various properties of silica.
We also thanks Enago (www.enago.jp) for the English language review.
We received financial support from the Research Program of ``Five-star Alliance" in ``NJRC Mater.~\& Dev."

%%%%%%%%%%%%%%%%%%%%%%%%%%%%%%%%%%%%%%%%%%
%%%%%%%%%%%%%%%%%%%%%%%%%%%%%%%%%%%%%%%%%%
\appendix
\section{Effects of periodic boundary condition}
The effect of using periodic boundary conditions on atom motions is investigated in this appendix.
The periodic boundary condition is used in all MD simulations. This introduces erroneous reflections of atom motion at the boundaries. The length, $a$, of the supercell limits the lowest wavelength of phonons in phonon spectra. The absence of phonons with wavelength, $\lambda$, longer than $a$ can be looked upon as the formation of virtual potential acting on atoms, which eliminates these long-wavelength phonons.
This effect becomes severe when $a$ is small. Let us estimate the effect of cell size on atom motions. A virtual periodic potential, $V_{p}$, is introduced to reproduce this effect. A simple form is assumed as follows:
\begin{equation}
 V_{p}(x) = \frac{V_{0}}{2} \left[ 1-\cos\left( 2 \pi \frac{x}{a} \right) \right],
 \label{eq:reflect-pot}
 \end{equation}
We use this virtual potential as a real one near the equilibrium position, $x=0$. Around $x=0$, this potential has a harmonic form with the force constant, $f^{\ast}= (V_{0}/2) (2 \pi/a)^{2}$. This yields the lowest frequency of the used cell, $\omega = \sqrt{2f^{\ast}/M}$. The lowest frequency obtained in the current cell size $2 \times 2 \times 2$ for crystal $\alpha$-quartz was about 50 cm$^{-1}$. Using the mass of oxygen for $M$, we obtain $f^{\ast}=23 \ \mu$dyn/\AA \, thus a potential barrier of $V_{0}=0.37$ eV. This magnitude is sufficient to raise the melting temperature by a few thousands of degrees K.

%%%%%%%%%%%%%%%%%%%%%%%%%%%%%%%%%%%%%%%%%%
\bibliography{thermo-refs,glass-refs}

% \end{thebibliography}

%%%%%%%%%%%%%%%%%%%%%%%%%%%%%%%%%%%%%%%%%%
\end{document}